\documentstyle[emulateapj,flushrt,tighten]{article} 
\input{psfig.sty}
\def\kms{\,{\rm km\,s^{-1}}}

\def\msun{\,{\rm M_\odot}}
\def\mden{\,{\rm M_\odot\,Mpc^{-3}}}

\def\Lya{Ly$\alpha$}
\def\etal{{et al.\ }}

\def\rei{{\rm rei}}

\def\HI{\hbox{H~$\scriptstyle\rm I\ $}}

\def\HII{\hbox{H~$\scriptstyle\rm II\ $}}
\def\HeI{\hbox{He~$\scriptstyle\rm I\ $}}
\def\HeII{\hbox{He~$\scriptstyle\rm II\ $}}
\def\bHeII{\hbox{He~$\scriptstyle\rm II$}}
\def\HeIII{\hbox{He~$\scriptstyle\rm III\ $}}

\newcommand\beq{\begin{equation}}
\newcommand\eeq{\end{equation}}
\newcommand{\ba}{\begin{eqnarray}}
\newcommand{\ea}{\end{eqnarray}}

\def\spose#1{\hbox to 0pt{#1\hss}}
\def\lta{\mathrel{\spose{\lower 3pt\hbox{$\mathchar"218$}}
      \raise 2.0pt\hbox{$\mathchar"13C$}}}
\def\gta{\mathrel{\spose{\lower 3pt\hbox{$\mathchar"218$}}
      \raise 2.0pt\hbox{$\mathchar"13E$}}}

\lefthead{Madau, Rees, Volonteri, Haardt, \& Oh}
\righthead{Early reionization by miniquasars}
\makeatletter

\newenvironment{figurehere}
  {\def\@captype{figure}}
  {}
\makeatother

\begin{document}
\submitted{}

\title{Early reionization by miniquasars}

\author{P. Madau\altaffilmark{1,2}, M. J. Rees \altaffilmark{3}, 
M. Volonteri\altaffilmark{1}, F. Haardt\altaffilmark{4}, \&
S. P. Oh\altaffilmark{5}}

\altaffiltext{1}{Department of Astronomy \& Astrophysics, University of 
California, Santa Cruz, CA 95064.}
\altaffiltext{2}{The Observatories of the Carnegie Institution of 
Washington, 813 Santa Barbara Street, Pasadena, CA 91101.}
\altaffiltext{3}{Institute of Astronomy, Madingley Road, Cambridge 
CB3 0HA, UK.}
\altaffiltext{4}{Dipartimento di Scienze, Universit\`a dell'Insubria/Sede 
di Como, Italy.} 
\altaffiltext{5}{Department of Physics, University of California, Santa 
Barbara, CA 93106.} 

\slugcomment{submitted to the ApJ}

\begin{abstract}
Motivated by the recent detection by the {\it Wilkinson Microwave Anisotropy 
Probe} of a large optical depth to Thomson scattering -- implying a very 
early reionization epoch -- we assess a scenario where the universe was
reionized by `miniquasars' powered by intermediate-mass black holes 
(IMBHs), the remnants of the first generation of massive stars. Pregalactic 
IMBHs form within minihalos above the cosmological Jeans mass collapsing at $z>20$, 
get incorporated through mergers into larger and larger systems, 
sink to the center owing to dynamical friction, and accrete cold material.
The merger history of dark halos
and associated IMBHs is followed by Monte Carlo realizations of the merger 
hierarchy in a $\Lambda$CDM cosmology. Our model is based on the 
assumptions that quasar activity is 
driven by major mergers and nuclear IMBHs accrete at the Eddington rate a
fraction of the gas in the merger remnant. The long dynamical frictional 
timescales leave many IMBHs `wandering' in galaxy halos after a minor 
merger. While seed IMBHs that
are as rare as the 3.5$\sigma$ peaks of the primordial density field evolve 
largely in isolation, a significant number of black 
hole binary systems will form if IMBHs populate the more numerous 
3$\sigma$ peaks instead. In the case of rapid binary coalescence a 
fraction of IMBHs will be displaced from galaxy centers and 
ejected into the IGM by the `gravitational rocket' effect, rather 
than accrete and shine as miniquasars. We show that, under 
a number of plausible assumptions for the amount of gas accreted onto 
IMBHs and their emission spectrum, miniquasars powered by IMBHs -- and 
not their stellar progenitors -- may be responsible
for cosmological reionization at $z\sim 15$. Reionization by miniquasars with
a hard spectrum may be more `economical' than stellar reionization, as 
soft X-rays escape more easily from the dense sites of star formation
and travel further than EUV radiation. Energetic photons will permeate the 
universe more uniformly, make the low-density diffuse IGM warm and weakly 
ionized prior to the epoch of reionization breakthrough, set an entropy 
floor, and reduce gas clumping. Future 21 cm observations may detect a 
preheated, weakly-ionized IGM in emission against the cosmic microwave background.
\end{abstract}

\section{Introduction}

In popular cold dark matter (CDM) hierarchical cosmogonies, the very 
first generation of metal-free (`Population III') stars is expected to
form in dark matter `minihalos' of total mass $M_h\gta 5\times 10^5\msun$ 
(e.g. Fuller \& Couchman 2000; Yoshida \etal 2003) condensing due to 
H$_2$ cooling from the high-$\sigma$ peaks 
of the primordial density field at redshift $z=20-30$. Recent numerical 
simulations of the collapse and fragmentation of primordial molecular 
clouds suggest that the first stars were predominantly very massive, 
$m_*\gta 100\,\msun$, a scale linked to H$_2$ chemistry and cooling 
(Bromm, Coppi, \& Larson 2001; Abel, Bryan, \& Norman 2002). 
Zero-metallicity very massive stars have spectra similar to that of a $\sim 
10^5\,$K blackbody (Tumlinson \& Shull 2000), and emit about 20 times more
Lyman-continuum photons per stellar baryon than a standard stellar 
population (Schaerer 2002; Bromm, Kudritzki, \& Loeb 2001).
A `top-heavy' primordial initial mass function (IMF), quite different 
from the 
present-day Galactic case, may have then played a crucial role in 
determining the ionization, thermal, and chemical enrichment history of the 
intergalactic medium (IGM) at early times.

The IGM is known to be highly ionized at least out to redshift $\sim 5.5$. 
While the excess \HI absorption measured in the spectra of $z\sim 6$ 
quasars in the Sloan Digital Sky Survey (SDSS) has been interpreted as the 
signature of the trailing edge of the cosmic reionization epoch (Djorgovski
\etal 2001; Fan \etal 2002; White \etal 2003), the recent detection by the 
{\it Wilkinson Microwave Anisotropy Probe} ({\it WMAP}) satellite of a 
large optical depth to Thomson scattering, $\tau_e=0.17\pm 0.04$ (68\%), 
suggests that the universe was reionized at much higher redshift, 
$z_{\rm rei}=17\pm3$ (assuming instantaneous hydrogen reionization, see 
Kogut \etal 2003; Spergel \etal 2003). This is an indication 
of significant star-formation activity at very early times. In the 
wake of {\it WMAP} remarkable results, several theoretical studies 
of stellar reionization scenarios have pointed out that an early generation of 
massive, metal-free stars with a top-heavy IMF may be necessary to produce 
a Thomson optical depth as high as the central value determined by {\it 
WMAP} (Cen 2003a; Ciardi, Ferrara, \& White 2003; Haiman \& Holder 2003; 
Sokasian \etal 2003; Wyithe \& Loeb 2003; Somerville \& Livio 2003).

Since, at zero metallicity, mass loss through radiatively-driven stellar 
winds or nuclear-powered stellar pulsations is expected to be negligible 
(Kudritzki 2002; Baraffe, Heger, \& Woosley 2001), Population  III stars 
will likely die losing only a small fraction of their mass. Nonrotating 
very massive stars in the mass window $140\lta m_*\lta 260\,\msun$ will disappear 
as pair-instability 
supernovae (Bond, Arnett, \& Carr 1984), leaving no compact remnants and 
polluting the universe with the first heavy elements (e.g. Schneider 
\etal 2002; Oh \etal 2001; Wasserburg \& Qian 2000). 
Stars with $40<m_*<140\,\msun$ and $m_*>260\,\msun$ are predicted instead 
to collapse to black holes (BHs) with masses exceeding half of the initial
stellar mass (Heger \& Woosley 2002). Barring any fine tuning
of the IMF of Population III stars, intermediate-mass black holes 
(IMBHs) -- with masses
above the 4--18$\,\msun$ range of known `stellar-mass' BHs (e.g. McClintock 
\& Remillard 2003) -- may then be the inevitable endproduct of the first 
episodes of pregalactic star formation.
Since they form in high-$\sigma$ density fluctuations,  
`relic' IMBHs and their descendants will tend to cluster in the bulges of
present-day galaxies as their host halos aggregate into more massive 
systems (Madau \& Rees 2001, hereafter Paper I; Islam, Taylor, \& Silk 
2003). The first IMBHs
may also have seeded the hierarchical assembly of the supermassive variety 
(SMBHs) observed at the center of luminous galaxies (Volonteri, Haardt, 
\& Madau 2003, hereafter Paper II; Volonteri, Madau, \& Haardt 2003, 
hereafter Paper III). 

Population III IMBHs accreting gas from the surrounding medium will shine 
as `miniquasars' at $z\sim 15$ and generate a soft X-ray background
that may catalyze the formation of H$_2$ molecules in dense regions and 
counteract the destruction by UV Lyman-Werner radiation (Cen 2003b; Glover 
\& Brandt 2003; Haiman, Abel, \& Rees 2000). The net effect would be an 
increase in the cooling rate and star formation efficiency of minihalos 
(but see Machacek, Bryan, \& Abel 2003 who find the positive feedback 
effect of X-rays to be quite mild). In this paper we point out that 
miniquasars represent an additional source of Lyman-continuum photons 
that must be taken into account  
in models of early reionization by Population III objects. Thin disk accretion
onto a Schwarzschild black hole releases about 100 MeV per baryon. If 
`seed' IMBHs were able to (say) double their initial mass via gas accretion, 
and just a few percent of the radiated energy were emitted above and close to 
the hydrogen Lyman edge, then {\it miniquasars would be more efficient at 
photoionizing the universe than their metal-free stellar progenitors.} 

While the basic idea is simply stated, it is not easy to quantify 
and model in detail. A minihalo at $z\sim 20$ will typically 
undergo several major mergers in a Hubble time. Pregalactic IMBHs 
will follow the merger history of their hosts, and may be able 
to accrete efficiently only in the inner densest regions of the merger 
remnant. Owing to the long dynamical frictional timescales, many IMBHs 
will not sink to the center and will be left `wandering' in galaxy halos 
after a minor merger. If seed holes are numerous enough, binary systems
may form in significant numbers. In this case, rather than accrete and 
shine as miniquasars, a fraction of IMBHs may be displaced from galaxy centers 
and 
ejected into the IGM by the `gravitational rocket' effect (Redmount 
\& Rees 1989) or by triple BH interactions (Paper II). Miniquasars are 
expected to be copious sources of soft X-ray photons, which will permeate the 
IGM more uniformly than possible with extreme ultraviolet (EUV, 
$\ge 13.6\,$eV) radiation (Oh 2000, 2001) and make it warm and 
weakly ionized prior to the epoch of reionization breakthrough (Venkatesan, Giroux, \& 
Shull 2001).

In this paper we make a first attempt at treating the growth and impact 
on the very early IGM of Population III IMBHs in the context of hierarchical 
structure formation theories, 
incorporating some of the essential astrophysics into a scenario  
for the reheating and reionization of the universe by miniquasars.
The plan is as follows. In \S~2 we outline a model
in which seed holes populate the rare 3.5$\sigma$ peaks of the primordial
density field. We show that, if cold material can be accreted efficiently 
onto IMBHs hosted in minihalos, then miniquasars may be responsible for 
cosmological reionization at $z\sim 15$. In \S~3 we present results for a 
case in which seed holes are more numerous but gas accretion occurs 
less efficiently, and address the dynamics of IMBH binary systems in 
shallow potential wells. In \S~4 we discuss the implications of an early 
reionization and reheating epoch by hard radiation from miniquasars. 
Finally, we summarize our results in \S~5. 
Unless otherwise stated, all results shown below refer to the currently 
favoured $\Lambda$CDM world model with $\Omega_M=0.3$, $\Omega_\Lambda=0.7$, 
$h=0.7$, $\Omega_b=0.045$, $\sigma_8=0.93$, and $n=1$.
 
\section{Build up of IMBH\lowercase{s} in merging minihalos}

A scenario for the birth, growth, and evolution of IMBHs that 
traces their formation in minihalos with virial temperatures $T_{\rm vir}<10^4\,$K
back to very high redshifts was put forward in Papers I and II. Seed IMBHs are 
placed far up in the dark halo merger tree, get incorporated 
into larger and larger halos, sink to the center owing to dynamical 
friction, and accrete a fraction of the gas in the merger remnant. The 
merger history of DM halos and associated IMBHs is followed through 
Monte Carlo realizations (based on 
the extended Press-Schechter formalism) of the merger hierarchy from early 
times to the present.

Here we improve on several aspects and adapt some prescriptions in
our model to the conditions expected during the epoch of first light.
Pregalactic seed IMBHs form within the mass ranges $20<m_\bullet<70\,\msun$
and $130<m_\bullet<200\,\msun$, as remnants of the first generation of 
massive metal-free stars that do not disappear as pair-instability 
supernovae.   
For simplicity, within these intervals the differential black hole mass 
function is assumed to be flat, i.e. the IMF-averaged BH mass is
115$\,\msun$. They form in isolation within minihalos 
above the cosmological Jeans mass collapsing at $z=24$
from rare $>\nu$-$\sigma$ peaks of the 
primordial density field. As our fiducial model we take $\nu=3.5$, 
corresponding in the assumed $\Lambda$CDM cosmology to minihalos of 
mass $M_{\rm seed}=1.3\times 10^6\,\msun$. As halos 
more massive than the $\nu$-$\sigma$ peaks 
contain a fraction erfc($\nu/\sqrt{2}$) ($=0.00047$ for $\nu=3.5$) of 
the mass of the universe, the mass density parameter of 
our `$3.5\sigma$' pregalactic holes is
\beq
\Omega_\bullet={0.00047\,\Omega_M\,\langle m_\bullet\rangle\over 
M_{\rm seed}}=10^{-7.9}. \label{ombh}
\eeq
This is 0.6\% of the mass density of the supermassive variety 
found in the nuclei of most nearby galaxies, $\Omega_{\rm SMBH}=(2.1\pm
0.3) \times 10^{-6}$ (Yu \& Tremaine 2002).  
We also study  a case in which seed holes are more numerous
and populate all peaks above $3\sigma$. Then 
erfc($3/\sqrt{2}$)=0.0027, $M_{\rm seed}=1.2\times 10^5\,\msun$, 
and the initial mass density of `$3\sigma$' IMBHs is 
$\Omega_\bullet=10^{-6.1}$, less than 40\% of the mass density of SMBHs today.
We do not consider lighter, more common halos than those originating from 
$3\sigma$ fluctuations, as they would fall below the minimum mass threshold for 
baryonic condensation via molecular hydrogen cooling, $M_{\rm TH}=1.25\times 10^5\,\msun$ according
to the simulations of Machacek, Bryan, \& Abel (2001). Seed holes that were much rarer
than the $3.5\sigma$ density  peaks would have little impact on the reionization of the IGM
at $z\gta 15$. Also, the assumed `bias' assures that almost all halos above $10^{11}\,\msun$ 
actually host a BH at later epochs. For simplicity, we start our calculations by placing all seed
IMBHs in their host halos at $z=24$, the highest redshift we are able to follow the merger hierarchy
to because of computational costs. More realistically, IMBHs are expected to form continuously
from Population III massive stars over a range of redshifts. 
 
Each halo is treated as a singular isothermal sphere (SIS) with density
$\rho\propto r^{-2},$ truncated at the virial radius. 
During the merger of two halos, the `satellite' progenitor with initial 
mass $M_s$ -- and its IMBH if it hosts one -- spirals in toward the 
center of the more massive pre-exisiting system (with mass $M_h>M_s$) on 
the dynamical friction (against the DM background) timescale. 
This depends on the orbital parameters of the satellite (van den Bosch
\etal 1999; Ghigna \etal 1998) and on tidal mass loss/evaporation. N-body
simulations show that massive satellites with mass ratios $P\equiv 
M_s/M_h>0.1$ (`major' mergers) sink rapidly without significant 
mass loss, while the lightest satellites ($P<0.01$) are almost unaffected 
by dynamical friction and do not suffer orbital decay (Taffoni \etal 2003).
In our calculations we adopt the expression for the dynamical friction 
timescale of a 
mass-losing satellite in a main halo suggested by Colpi, Mayer, \& Governato
(1999).

\begin{figurehere}
\vspace{+0.2cm}
\centerline{
\psfig{file=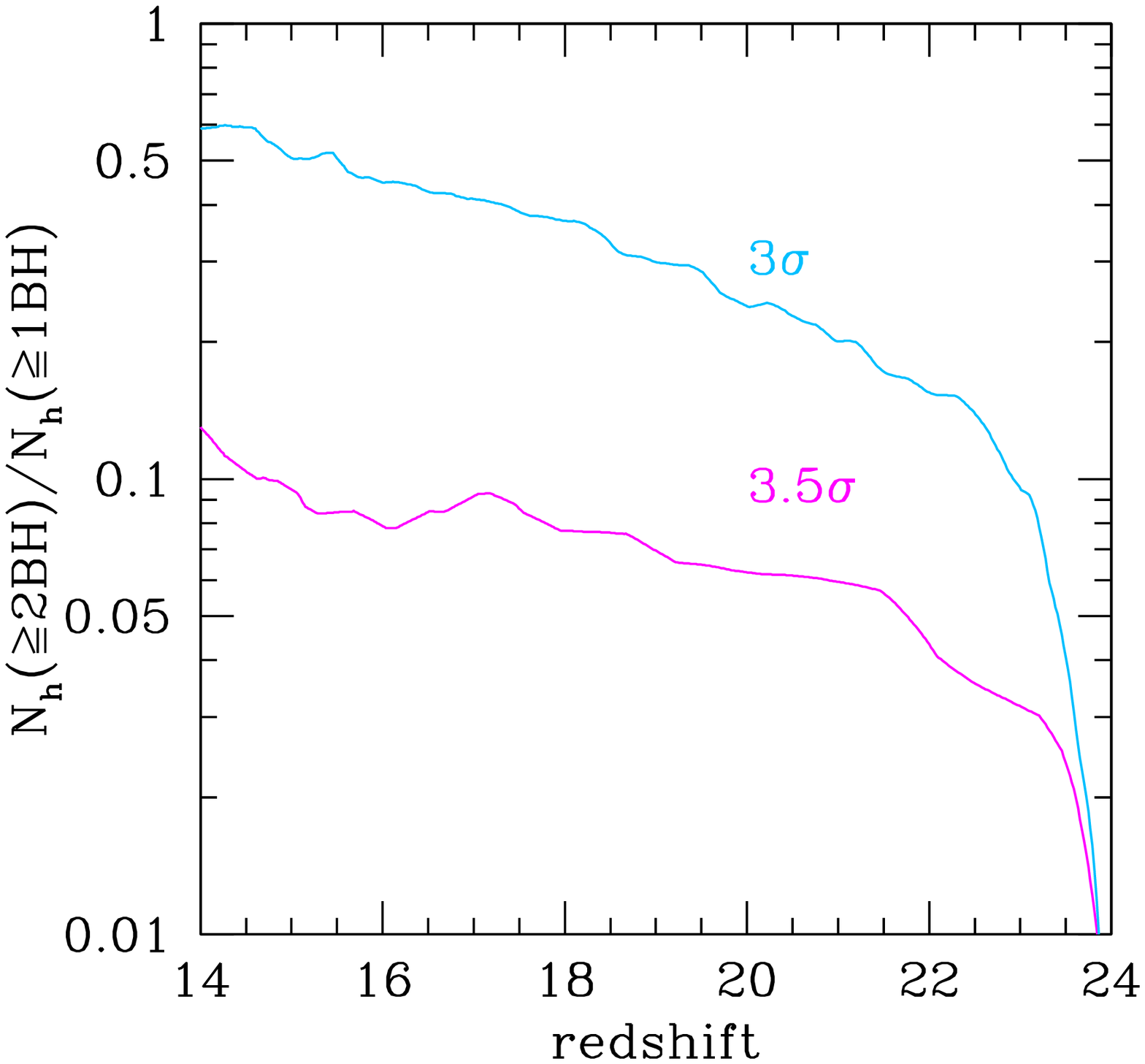,width=3.0in}}
\caption{\footnotesize Fraction of BH host halos that contains two or more
seed IMBHs as a function of redshift, averaged over 50 `trees'. Two 
curves are shown, one assuming BHs
form in isolation within minihalos collapsing at $z=24$ from 3.5$\sigma$ 
peaks, the other from 3$\sigma$ peaks instead. For illustrative 
purposes, dynamical 
interactions between BHs have been switched off along cosmic history.
}
\label{fig1}
\vspace{+0.5cm}
\end{figurehere}

In our fiducial model ($\nu=3.5$), the density of seed holes is small enough 
that the merging of two minihalos both hosting a BH is a rare event, and IMBHs
evolve largely in isolation. This is shown in Figure \ref{fig1}, where
the number of halos containing two or more primordial IMBHs,  normalized to 
the total number of BH hosts, is plotted as a function of redshift. 
The ratio, as expected, increases with time and tends to saturate at lower 
redshifts. For
illustrative purposes, the realizations in this figure have been run 
neglecting any dynamical interaction between BHs along cosmic history. 
At the very high redshifts of interest here this is a good 
assumption for $\nu=3.5$ as the fraction of multiple systems is typically
less than 10\%. 
If seed holes populate the $\nu=3$ peaks instead, the fraction of multiple
BH hosts can exceed 50\% and bound binaries form in 
significant numbers. We will discuss the dynamics of IMBH binaries in \S~3.       
\subsection{Accretion onto IMBH\lowercase{s}}

Hydrodynamic simulations of major mergers have shown that a significant 
fraction of the gas in interacting galaxies falls to the center of the 
merged system (Mihos \& Hernquist 1996): the cold gas may be eventually 
driven into the very inner regions, fueling an accretion episode and the 
growth of nuclear BHs (Kauffmann \& Haehnelt 2000; Cavaliere \& Vittorini 
2000). Since the local SMBH mass density is consistent
with the integrated luminosity density of quasars (Yu \& Tremaine 2002; 
Fabian 2003), the fraction of cold gas ending up in the hole must 
depend on the properties of the host halo in such a way to ultimately 
lead to the observed correlation between stellar velocity dispersion 
and SMBH mass (Ferrarese \& Merritt 2000; Gebhardt \etal 2000).  
In Paper II it was assumed that in each major merger the hole in the most
massive halo accretes at the Eddington rate a gas mass that scales with 
the fifth power of the circular velocity of the host halo: the normalization 
was fixed a posteriori in order to reproduce the observed local 
$m_{\rm BH}-\sigma_*$ relation (Ferrarese 2002). Our simple model was 
shown to explain remarkably well the observed luminosity 
function of optically-selected quasars in the redshift range $1<z<5$. 
Accretion was inhibited onto all BHs hosted in halos with virial 
temperature $T_{\rm vir}\lta 10^4\,$K, as below this temperature atomic 
cooling is not effective in allowing the gas to sink to the center. 

Prior to the epoch of reionization, however, cold material may be 
efficiently accreted onto IMBHs hosted in minihalos. As mentioned 
in \S~1, gas condensation in the first
baryonic objects is possible through the formation of H$_2$ molecules,
which cool efficiently via roto-vibrational transitions even at virial 
temperatures of a few hundred kelvins. In the absence of a UV 
photodissociating flux and of ionizing X-ray radiation, three-dimensional
simulations of early structure formation show that the fraction of 
cold, dense gas available for accretion onto IMBHs or star formation     
exceeds 20\% above $M_h=10^6\,\msun$ (Machacek, Bryan, \& Abel 2003).
And while radiative feedback (photodissociation and photoionization) 
from the 
progenitor massive star may initially quench BH accretion within the original
host minihalo, {\it new cold material will be readily available through the
hierarchical merging of small gaseous subunits}.  
In this section we will assume that gas accretion onto BHs can actually 
occur with high efficiency in minihalos just above the cosmological Jeans 
mass. 

As there is no compelling reason to expect the $m_{\rm BH}-\sigma_*$ 
relation to be 
set in primordial structures and be satisfied at the very high redshifts 
$(z\sim 15$) considered in this work, we run two different sets of 
realizations. In the first we assume that in each major merger the hole 
in the main halo {\it triples} its mass via gas accretion, 
$\Delta m_{\rm acc}=2m_{\rm BH}$, while in the second the hole accretes 
a fixed fraction of the host halo mass, $\Delta m_{\rm acc}=10^{-3}\,M_h$.
In both realizations this mass is added to the hole at the Eddington rate,
$\dot m_E=4\pi Gm_pm_{\rm BH}/(c\sigma_T\epsilon)$, for a radiative 
efficiency $\epsilon=10\%$. In the case of a merger in which only the
satellite halo contains a BH, the hole is placed at the center of the merger
remnant after one dynamical friction timescale, and must await a subsequent 
major merger event with a (smaller) satellite to start accreting. Each accretion 
episode begins after 
one dynamical timescale (measured at radius $0.1\,r_{\rm vir}$) and typically
lasts for a few Salpeter timescales, $t_S=m_{\rm BH}/\dot m_E=4.5\times 
10^7\,{\rm yr}~(\epsilon/0.1)^{-1}$,    
until a mass $\Delta m_{\rm acc}$ has been added to the hole. 
Figure 2 shows how the comoving mass density of IMBHs builds up 
with cosmic time: with the adopted prescriptions, by a redshift of 14 this 
never exceeds 3\% of the mass density of SMBHs inferred by Yu \& Tremaine 
(2002) in local galaxies.

\begin{figurehere}
\vspace{+0.2cm}
\centerline{
\psfig{file=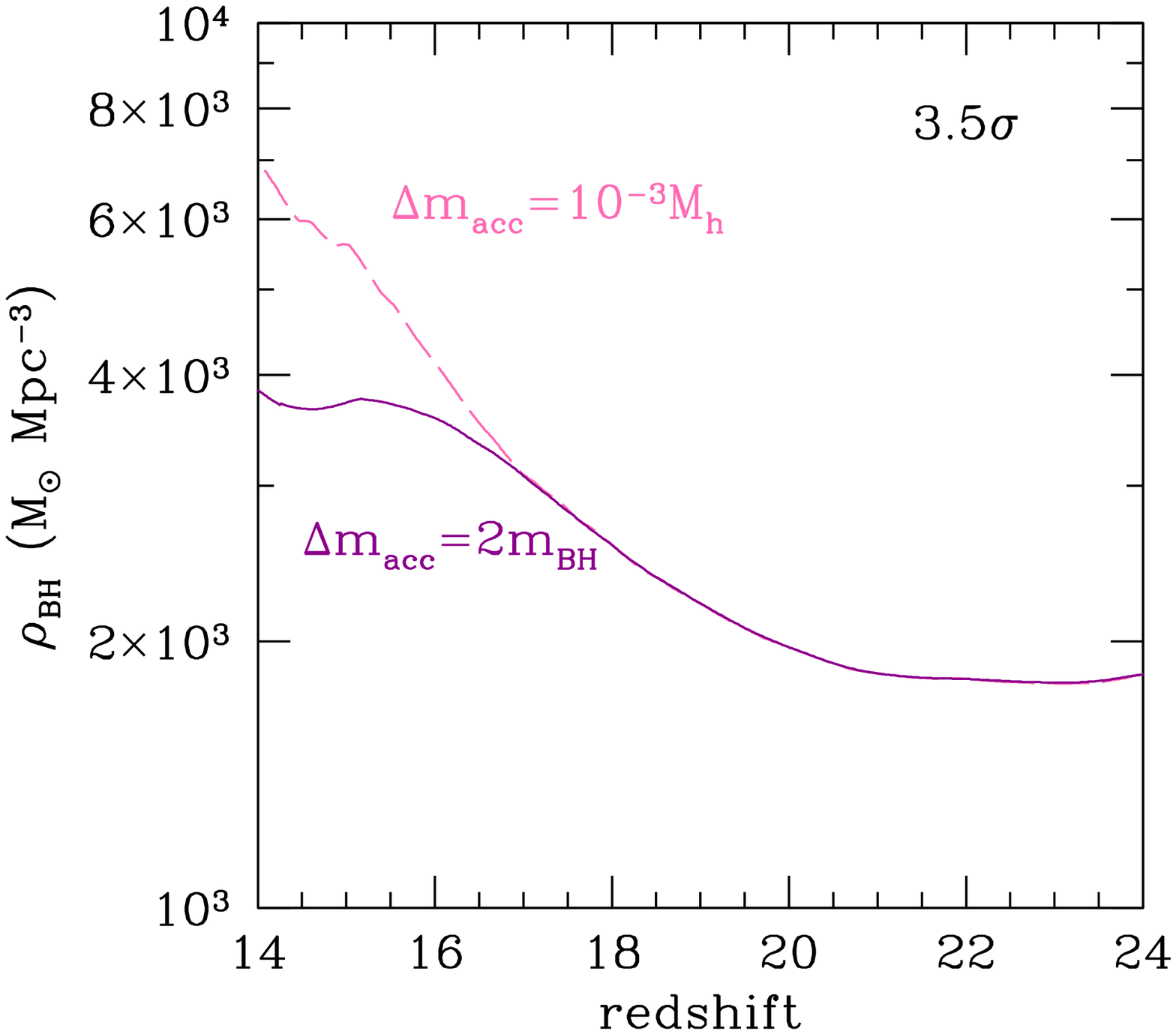,width=3.0in}}
\caption{\footnotesize Comoving mass density in IMBHs as a function of 
redshift 
for the two different accretion recipes discussed in the text. Gas accretion
onto BHs is assumed to occur in minihalos only after a major merger.
Seed IMBHs form in minihalos collapsing at $z=24$ from 3.5$\sigma$ density 
peaks.
}
\label{fig2}
\end{figurehere}

\subsection{Miniquasars}

Assume now that a fraction $f_{\rm UV}$ of the bolometric power radiated
by our Population III miniquasars is emitted as hydrogen-ionizing photons 
with mean energy $\langle h\nu\rangle$. For a given bolometric emissivity,
the number of ionizing photons scales as $f_{\rm UV}/\langle h\nu\rangle$. 
One of the biggest uncertainties in discussing early reionization by 
miniquasars is their unknown emission spectrum. If the shape of the emitted 
spectrum from miniquasars followed the mean spectral energy distribution 
of the quasar sample in Elvis \etal (1994), then $f_{\rm UV}=0.3$ 
and $\langle h\nu\rangle=3\,$ryd, hence $f_{\rm UV}/\langle h\nu\rangle=0.1\,$
ryd$^{-1}$. Miniquasars powered by IMBHs, however, are likely to be harder emitters 
than quasars. The hottest blackbody temperature 
in a Keplerian disk damping material onto a BH at the Eddington rate is
$kT_{\rm max}\sim 1\,{\rm keV}\,m_{\rm BH}^{-1/4}$ (Shakura \& Sunyaev 1973),
where the hole mass is measured in solar mass units. The characteristic 
multicolor (cold) disk spectrum follows a power-law with $L_\nu\sim 
\nu^{1/3}$ at $h\nu<kT_{\rm max}$. Close to the
epoch of reionization at $z\sim 15$, most IMBHs holes will have masses in the 
range $200-1000\,\msun$ (Fig. \ref{fig3}), hence $kT_{\rm max}\sim 
0.2-0.3\,$keV.
The spectra of `ultraluminous' X-ray sources (ULXs, Colbert \& Mushotzky 
1999) in nearby galaxies actually require both a soft and a hard 
component of comparable luminosities to describe the continuum 
emission. While the soft components are well fit by cool multicolor 
disk blackbodies
with $kT_{\rm max}\simeq 0.15\,$keV, which may indicate IMBHs (e.g.
Miller \etal 2003), the nonthermal power-law component has spectral slope 
$L_\nu\propto \nu^{-\alpha}$, with $\alpha\approx 1$. A power-law with 
slope $\alpha=1$ extending from 2 keV down to 13.6 eV has $f_{\rm UV}=1$
and $\langle h\nu\rangle=5\,$ryd, hence $f_{\rm UV}/\langle h\nu\rangle=
0.2\,$ryd$^{-1}$. Figure \ref{fig4} shows the cumulative number of ionizing 
photons produced per hydrogen atom for different values of 
$f_{\rm UV}/\langle h\nu\rangle$. A few ionizing hard photons per atom 
(less than for a stellar spectrum because of secondary ionizations and  
larger mean free path, see \S~4) should suffice 
to reionize the universe and keep the gas in overdense regions and filaments 
photoionized against radiative recombinations. 
 
\begin{figurehere}
\vspace{+0.2cm}
\centerline{
\psfig{file=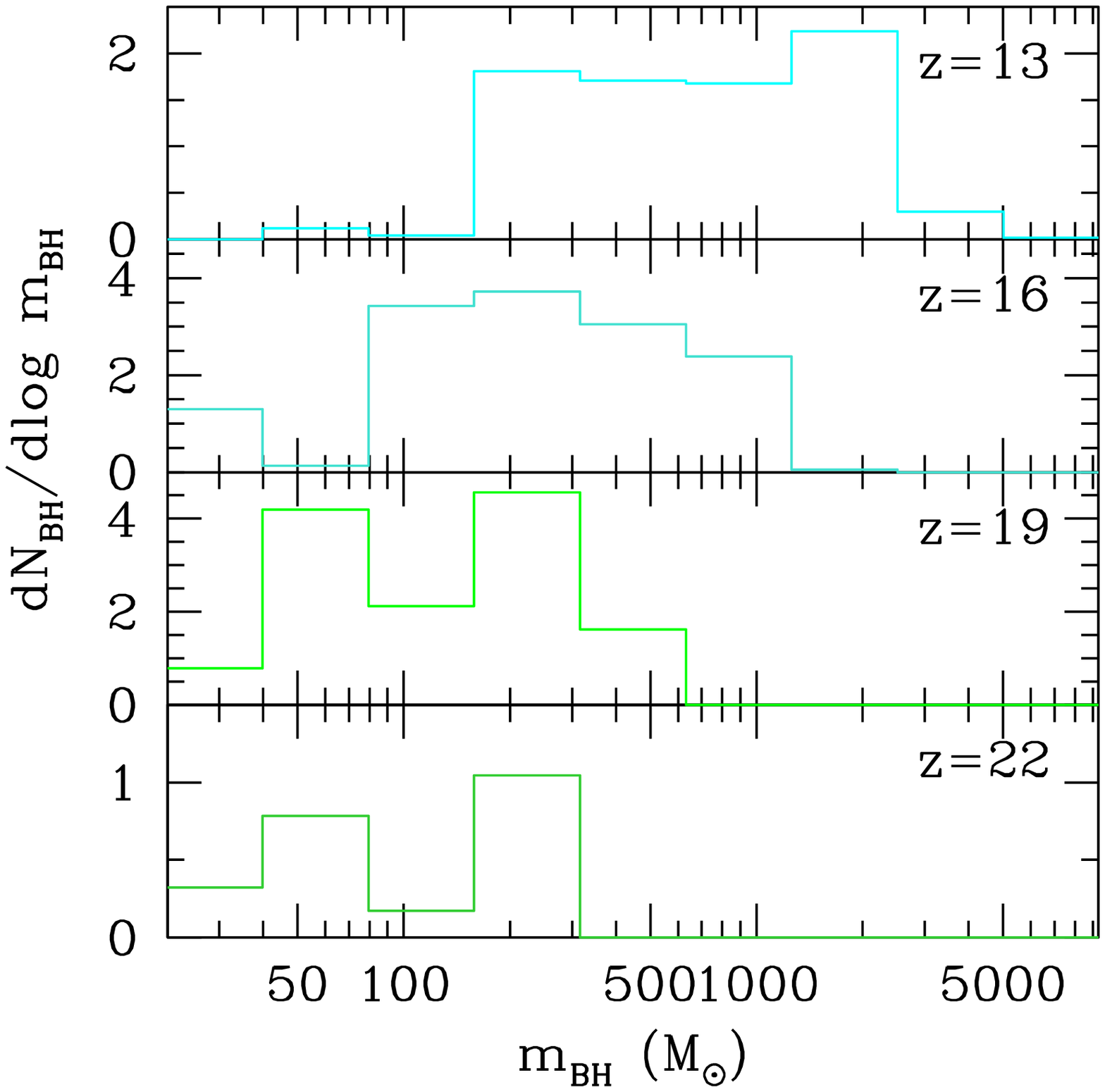,width=3.0in}}
\caption{\footnotesize Mass function of accreting IMBHs predicted at 
four different redshifts by our fiducial model. Units are arbitrary. 
All seed holes are assumed to form at $z=24$ from 3.5$\sigma$ density peaks.
}
\label{fig3}
\end{figurehere}

\begin{figurehere}
\vspace{+0.2cm}
\centerline{
\psfig{file=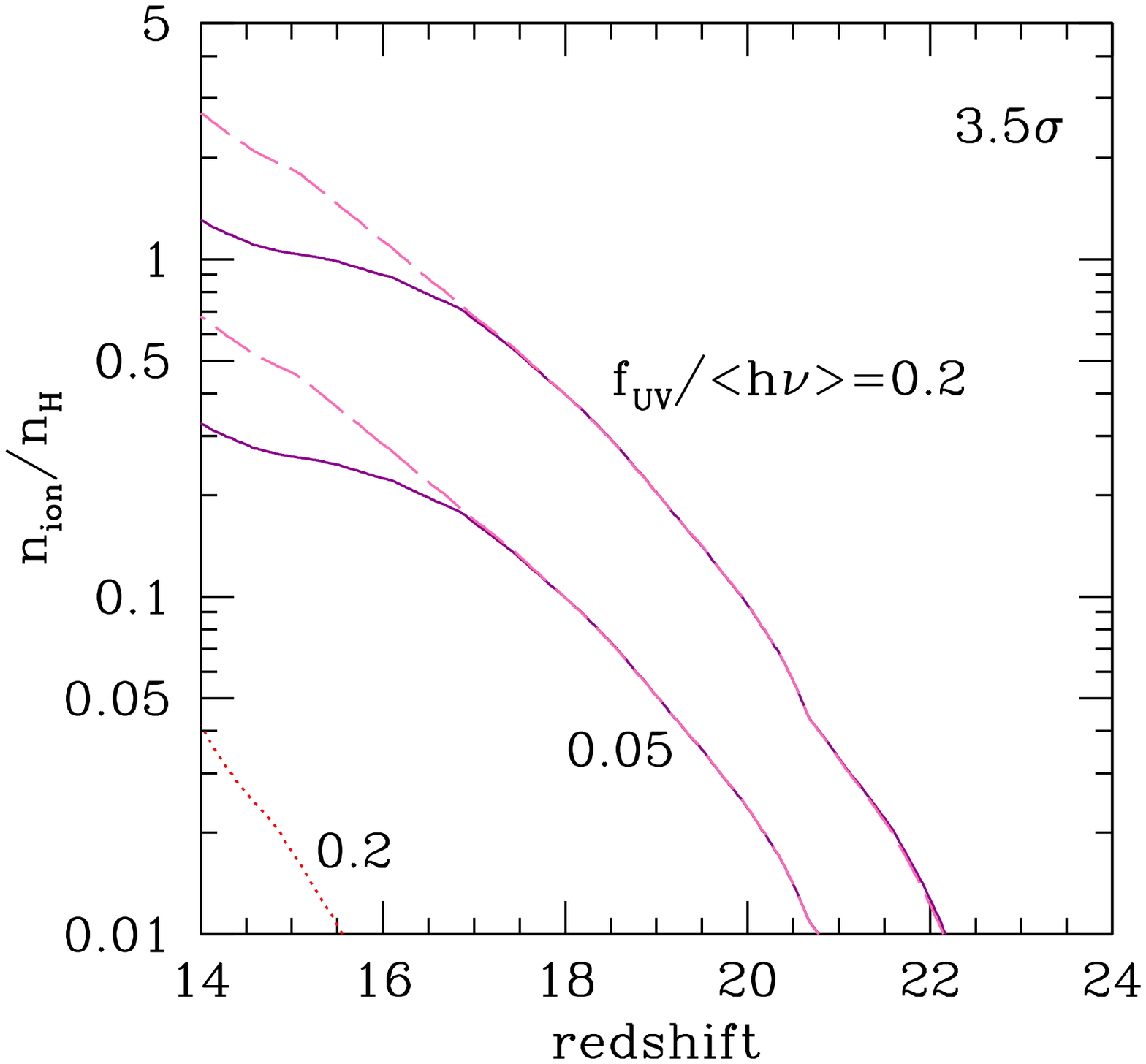,width=3.0in}}
\caption{\footnotesize Cumulative number of ionizing photons per 
hydrogen atom produced by miniquasars, for different values of the
fraction $f_{\rm UV}$ of the bolomeric power that is emitted as radiation 
above 1 ryd with mean energy $\langle h\nu\rangle$ (in ryd). 
{\it Solid curves:} in each major 
merger the BH in the main halo accretes a mass $\Delta m_{\rm acc}=
2m_{\rm BH}$. {\it Dashed curves:} same with $\Delta m_{\rm acc}=10^{-3}\,M_h$.
{\it Dotted curve:}  same with $\Delta m_{\rm acc}=10^{-3}\,M_h$, 
$f_{\rm UV}/\langle h\nu\rangle=0.2\,$ryd$^{-1}$,  
but with gas accretion suppressed in minihalos with virial 
temperature $T_{\rm vir}\lta 10^4\,$K.
}
\label{fig4}
\vspace{+0.3cm}
\end{figurehere}

To illustrate the implications of these results, consider the 
following estimate for the number of H-ionizing photons emitted by the 
initial population of progenitor massive stars. In our model the 
fraction, $f_*$, of cosmic baryons incorporated into Population III 
massive stars -- progenitors of seed IMBHs -- at $z=24$ is,
\beq
f_*={\langle m_*\rangle\over \langle m_\bullet\rangle}\,
{\Omega_\bullet\over \Omega_b}\approx 5\times 10^{-7}, 
\eeq
where $\Omega_b$ is the baryon density parameter. Zero-metallicity 
stars in the range
$40<m_*<500\,\msun$ emit about 70,000 photons above 1 ryd per stellar baryon 
(Schaerer 2002). The total number of ionizing photons produced per baryon by 
progenitor Population III stars is then, using equation (\ref{ombh}), 
\beq
{n_{\rm ion}\over n_b}\approx 70,000 f_*\approx 0.04, \label{ionb}
\eeq 
well below what is needed to reionize the universe. 
Figure \ref{fig4} shows that,
if $f_{\rm UV}/\langle h\nu\rangle$ is greater than $0.1\,$ryd$^{-1}$ 
and gas is accreted efficiently
onto IMBHs, then {\it miniquasars may be 
responsible for cosmological reionization at redshift $\sim 15$}. 
We will discuss in more details some of 
the implications of reheating by hard radiation from miniquasars in \S~4.

\section{Reduced accretion efficiency and dynamics of IMBH binaries}

Miniquasars powered by IMBHs forming in 3.5$\sigma$ peaks produce 
more ionizing photons than their progenitors, and will reionize the IGM 
by $z\sim 15$. Such conclusion is based on the assumption that 
IMBHs can grow rapidly via gas accretion during major mergers involving 
minihalos 
and shine in the EUV/soft X-rays. Local and global feedback effects, however, 
may act to reduce or even inhibit accretion in minihalos where 
atomic cooling is inefficient. More massive halos with $T_{\rm vir}>10^4\,$K, 
in which atomic cooling can operate, are expected to be very rare at these 
redshifts. If gas accretion was completely suppressed 
in minihalos with $T_{\rm vir}\lta 10^4\,$K, i.e. with masses
below $10^{7.9}\,[(1+z)/20]^{-3/2}\,\msun$, the lower emissivity 
would either shift the reionization epoch to lower redshift or make 
miniquasars a sub-dominant source of ionizing photons 
(Fig. \ref{fig4}), at least until halos 
massive enough that gas can initially cool and contract via 
excitation of hydrogen \Lya, collapsed in large numbers.

Still, a viable scenario for early reionization by miniquasars may 
be one in which  
gas accretion onto IMBHs is much reduced compared to our fiducial model in 
\S~2, but seed holes are more numerous at the start in order to sustain 
the early production of ionizing radiation. Black holes forming in 3$\sigma$ 
density fluctuations cannot be assumed to evolve in isolation, and the 
dynamics of BH binaries in shallow potential wells must be addressed.
In this section we shall assess a model in which IMBHs populate all halos 
more massive 
than $M_{\rm seed}=1.2\times 10^5\,\msun$, corresponding to the
$3\sigma$ peaks of the primordial density field. In each major 
merger a mass $\Delta m_{\rm acc}=10^{-5}\,M_h$ (two orders of
magnitude smaller than in our fiducial model) 
is now added to the hole in the more massive halo at the Eddington rate,
i.e. seed IMBHs do not grow appreciably due to gas accretion and 
miniquasars are typically `on' only for a small fraction of the Salpeter 
timescale. Note that a high accretion recipe with $\Delta m_{\rm acc}=
10^{-3}\,M_h$ (say) is not viable in the case of 3$\sigma$ seed holes, 
as it generates a $\rho_{\rm BH}$ that exceeds the mass density of SMBHs 
observed in nearby galaxies.   
  
\subsection{IMBH binaries}

As shown in Figure \ref{fig5}, if seed holes are as numerous as the 
3$\sigma$ density peaks, binary systems may form in significant numbers. 
The merging -- driven by dynamical friction against the DM -- of two 
minihalo$+$IMBH 
systems with mass ratio $P\gta 0.1$ will drag in the smaller hole towards 
the center, leading to the formation of a bound IMBH binary in the violently
relaxed core of the newly merged system. In massive galaxies at low redshift, 
the subsequent evolution of the binary may be largely determined by the 
central 
stellar distribution. The binary initially shrink (`hardens') by dynamical 
friction from distant stars acting on each BH individually, and later via
three-body interactions, i.e., by capturing the stars that pass
within a distance of the order of the binary semi-major axis and ejecting 
them at much higher velocities (Begelman, Blandford, \& Rees 1980).
Dark matter particles will be ejected by decaying binaries in the same way 
as the stars, i.e. through the gravitational slingshot. Their contribution to 
binary hardening will be approximately weighted by the fraction 
$f_{\rm DM}=\rho_{\rm DM}/(\rho_{\rm DM}+\rho_*)$, typically
$\ll 1$ in massive galaxy cores.

\begin{figurehere}
\vspace{+0.2cm}
\centerline{
\psfig{file=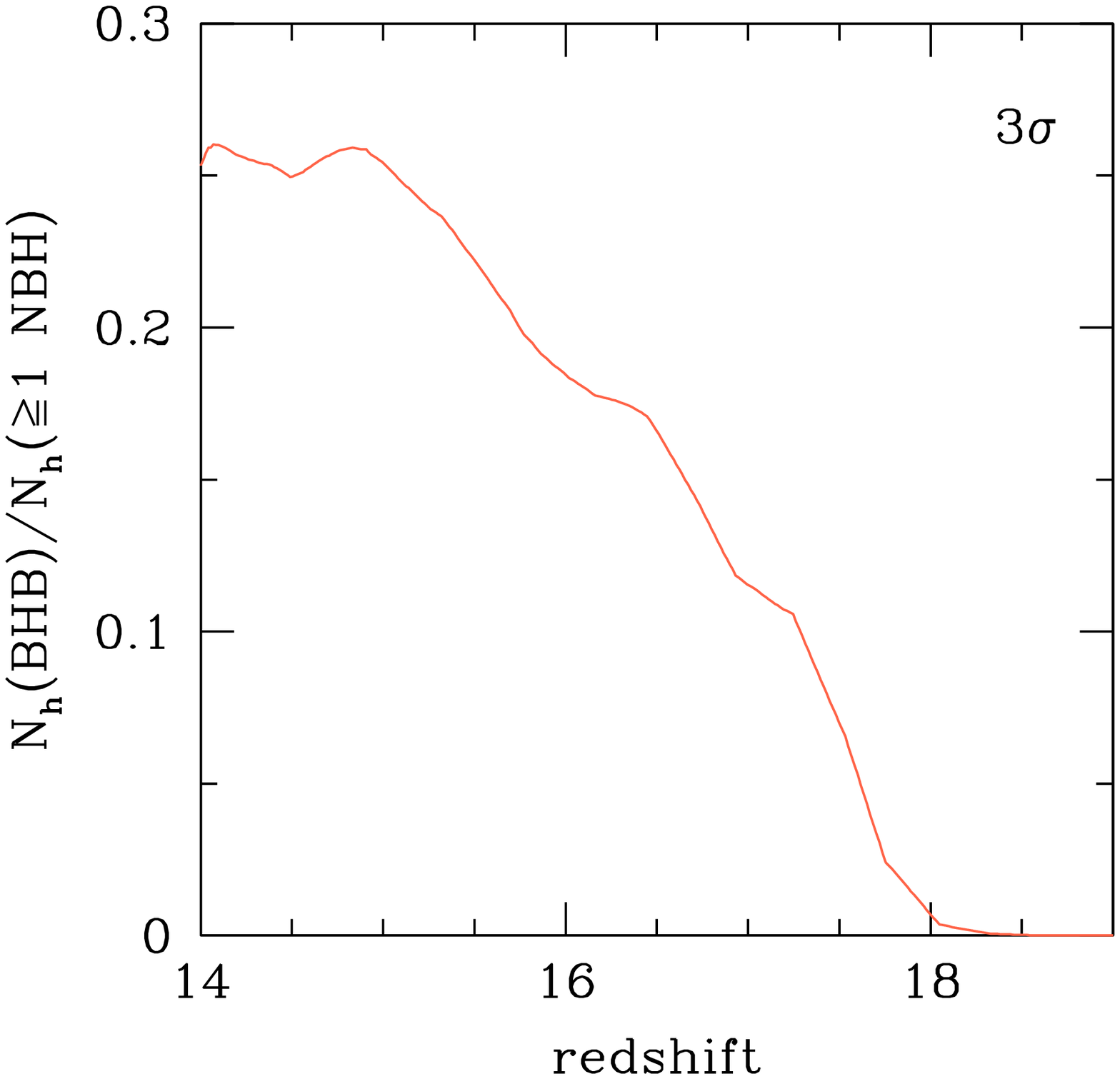,width=3.0in}}
\caption{\footnotesize Fraction of all halos hosting nuclear black holes 
(NBHs) that contain a hard BH binary (BHB) system, as a function 
of redshift. Binaries are `hard' if their separation is $a\le a_h$ (see 
text for details). Seed IMBHs form in minihalos collapsing 
at $z=24$ from 3$\sigma$ density peaks. In this figure we have 
assumed that black hole binaries do not shrink and 
coalesce (i.e. we set the hardening timescale $t_h\rightarrow \infty$).
}
\label{fig5}
\vspace{+0.5cm}
\end{figurehere}

In minihalos a numerous population of low-mass stars may be present 
if the IMF were bimodal, with a second peak at 1-2 $\msun$, as suggested by 
Nakamura \& Umemura (2001). Otherwise the binary will be losing orbital 
energy to the dark matter background. In the limit $f_{\rm DM}\gg 1$ 
(cf. Paper III), a binary with BH masses $m_1\ge m_2$ and semimajor axis 
$a(t)$ in an SIS dark halo of density $\rho$ and one-dimensional velocity 
dispersion $\sigma$, becomes `hard' when the binary separation falls 
below the value
\beq
a_h={Gm_2\over 4\sigma^2}=0.04\,{\rm pc}~\left({m_2\over 10^3\,\msun}\right)\,
\sigma_5^{-2}
\eeq
(Quinlan 1996), where $\sigma_5$ is measured in units of 
$5\,\kms$. The hardening of the binary 
modifies the density profile, removing mass 
interior to the binary orbit, depleting the galaxy core of dark matter, and slowing 
down further hardening. If ${\cal M}_{\rm ej}$ is the mass ejected 
by the BH pair, the binary evolution and its effect on galaxy cores are 
determined by two dimensionless quantities: the hardening rate
\beq
H={\sigma\over G\rho}{da^{-1}\over dt},
\label{eqH}
\eeq
and the mass ejection rate
\beq
J={1\over (m_1+m_2)}\,{d{\cal M}_{\rm ej}\over d\ln a^{-1}}.
\label{eqJ}
\eeq
The quantities $H$ and $J$ can be found from scattering experiments 
that treat the test particle-binary encounters one at a time (Quinlan 1996).
We assume as in Papers II and III that the removal of matter creates a 
core of radius $r_c$ and constant
density $\rho_c\equiv \rho(r_c)$, eroding a preexisting isothermal cusp
with $\rho\propto r^{-2}$. The total mass ejected as the binary shrinks 
from $a_h$ to $a$ is then ${\cal M}_{\rm ej}=4\sigma^2 r_c/3G$, while
the core radius grows as
\begin{equation}
r_c(t)={3\over4 \sigma^2}G(m_1+m_2)\int_{a(t)}^{a_h}\,J(a)da/a.
\label{rc}
\end{equation}
The binary separation quickly falls below $r_c$ and subsequent evolution
is slowed down due to the declining matter density, with a hardening time,
\beq
t_h=|a/\dot a|={2\pi r_c(t)^2\over H\sigma a},
\label{thard}
\eeq
that becomes increasingly long as the binary shrinks. If hardening continues 
down to a separation
\beq
a_{\rm gr}= 1\,{\rm AU}~~\left[{(m_1+m_2)m_1m_2\over 10^{9.3}\,\msun^3
}\right]^{1/4},
\eeq
the binary will coalesce within 200 Myr (the Hubble time at $z=18$) due 
to the emission of gravitational waves. 
The pair must then manage to shrink by a factor 
$a_h/a_{\rm gr}\sim 5000$ for gravity wave emission to become efficient.  
Since the hardening and mass ejection rate coefficient are 
$H\approx 15$ and $J\approx 1$ in the limit of a very hard binary (Quinlan 1996),
the hardening time in equation (\ref{thard}) can be rewritten (for $m_1=m_2$) as
\beq
t_h\approx 1.3\times 10^5\,{\rm yr}~\left({m_2\over 10^3\,\msun}\right)\,
\sigma_5^{-3}\,{a_h\over a}\,\ln^2(a_h/a). \label{lnth}
\eeq
This becomes longer than the then Hubble time already at $a_h/a\sim 
100$. Note that for an equal mass binary $a_h/a_{\rm gr}\propto
m_2^{1/4}\sigma^{-2}$. Thus even if the $m_{\rm BH}\propto \sigma^4$ 
relation (Tremaine \etal 2002) were to hold at very high redshifts, 
for gravitational radiation to induce coalescence a hard BH binary must 
decay by a larger factor in minihalos with small velocity 
dispersion than in more massive galaxies. At these early epochs, 
when the BH occupation
fraction is still small, we find that triple BH interactions are too 
rare to cause binaries to shrink (cf. Paper II). In the absence of any 
other mechanism for removing orbital angular momentum, the decay is then
expected to stall at a separation much greater than $a_{\rm gr}$. 
This is even more true as it is the total matter density that was allowed 
to decrease in equation (\ref{rc}), not the density of low-angular 
momentum particles that get close enough to extract energy from the binary, 
i.e. we have assumed the `loss cone' is constantly refilled. Note that
this would be a good assumption if orbital angular momentum 
losses to stars were the dominant mechanism of binary shrinkage, since the 
stellar relaxation timescale
\beq
t_r=\frac{0.34\sigma^3}{G^2 m_* \rho_c \ln\Lambda}\,=150\,{\rm Myr}
\left(\frac{r_c}{0.5\,\rm pc}\right)^2\left(\frac{4}
{\ln \Lambda}\right)\sigma_5,
\label{trelax}
\eeq
is typically much shorter than the hardening time in minihalos, and two-body 
scatterings will efficiently refill the loss cone.

\subsection{Gravitational rocket and the depletion of IMBHs from host 
minihalos}

The extent with which BH pairs loose angular momentum to DM particles 
or stars is one of the major uncertainties in 
computing merger timescales, and makes it difficult to construct viable 
merger scenarios for BH binaries (e.g. Milosavljevic \& Merritt 2001).
For illustrative purposes, we run here two different sets of realizations.
In the first the hardening is due to the ejection of DM particles, the
hardening timescale is long (from eqs. \ref{thard}-\ref{lnth}), 
and binaries stall. Another possibility is that gas processes, rather than 
three-body interactions with stars or DM, may induce IMBH binaries to shrink 
rapidly and coalesce (e.g. Armitage \& Natarajan 2002; Gould \& Rix 2000). 
Gas in a system containing a 
hard binary will fall inward in just the same way as if there were a single 
hole at the centre, as the binary separation is much smaller than the 
Bondi radius, $r_B=Gm_{\rm BH}/c_\infty^2=0.5\,{\rm pc}\,(m_{\rm BH}/1000\,
\msun)\,(T/1000\,{\rm K})^{-1}$. It will heat up and radiate away its 
binding energy. Some material will be accreted by each individual hole while 
some will be flung out, removing binary orbital angular momentum and
hastening the merger process. Further infall may replenish the gas reservoir 
on a dynamical timescale and keep the drag going. The details are unclear. 
As an extreme case we have therefore run a second 
set of realizations in which once a hard binary forms with separation 
$a_h$, it coalesces {\it instantaneously}. 
In the shallow 
potential wells of minihalos, the growth of IMBHs will then be halted by the
`gravitational rocket', the recoil due to the non-zero net linear momentum 
carried away by gravitational waves in the coalescence of two unequal 
mass black holes. Radiation recoil is a strong field effect that depends on 
the lack of 
symmetry in the system, and may eject IMBHs from the cores of minihalos. To 
date,
the outcome of a gravitational rocket remains uncertain, as fully general 
relativistic
numerical computations of radiation reaction effects are not available at the 
moment. The predicted recoil velocity, $v_{\rm CM}$, may be bracketed  
(Brandt \& Anninos 1999; Wiseman 1992) on the one side by quasi-Newtonian 
calculations predicting $v_{\rm CM}<100\,\kms$ (Fitchett 1983), and on 
the other by extrapolated perturbative results on test particles (Fitchett \& 
Detweiler 1984), allowing for rocket velocities up to $v_{\rm CM}=700\,\kms$
for $m_1=1.2\,m_2$ in a Schwarzschild geometry.

According to quasi-Newtonian calculations for circular orbits, while the 
binary shrinks its center of mass spirals outward with a velocity
\beq
v_{\rm CM}=1500\,\kms \left[\frac{f(m_1/m_2)}{f_{\rm max}}\right]
\left(\frac {r_s}{r_{\rm ms}}\right)^{4},
\eeq
where $r_s=2G(m_1+m_2)/c^2$ is the Schwarzschild radius of the system, $r_{\rm 
ms}=6G m_1/c^2$ is the radius of the closest stable circular orbit for two 
non-rotating holes (Clark \& Eardley 1977), and the function 
\beq
f(m_1/m_2)=|(1+\frac{m_1}{m_2})^{-2}(1+\frac{m_2}{m_1})^{-3}(1-\frac{m_2}{m_1})|,
\eeq
reaches a maximum value $f_{\rm max}=0.01789$ for $m_1=2.6m_2$.
The extrapolated perturbative results (Fitchett \& Detweiler 1984) yield
considerably higher recoil velocities for $m_1<4\,m_2$, while for $m_1>7\,m_2$ 
both calculations converge to recoil velocities below $15\,\kms$.
We adopt here the extrapolated perturbative results, 
as this assumption can be the most disruptive and lead to the larger 
modifications in terms of IMBH occupation fraction. The coalesced binary 
will leave the galaxy altogether if its recoil velocity exceeds the escape 
speed of the halo, giving origin to a population of {\it intergalactic} 
IMBHs.\footnote{We have assumed that the final mass $m_{\rm BH}$ of the hole after 
coalescence satisfies the entropy-area relation (maximally efficient radiative 
merging): $m_{\rm BH}^2=m_1^2+m_2^2$.}

\begin{figurehere}
\vspace{+0.2cm}
\centerline{
\psfig{file=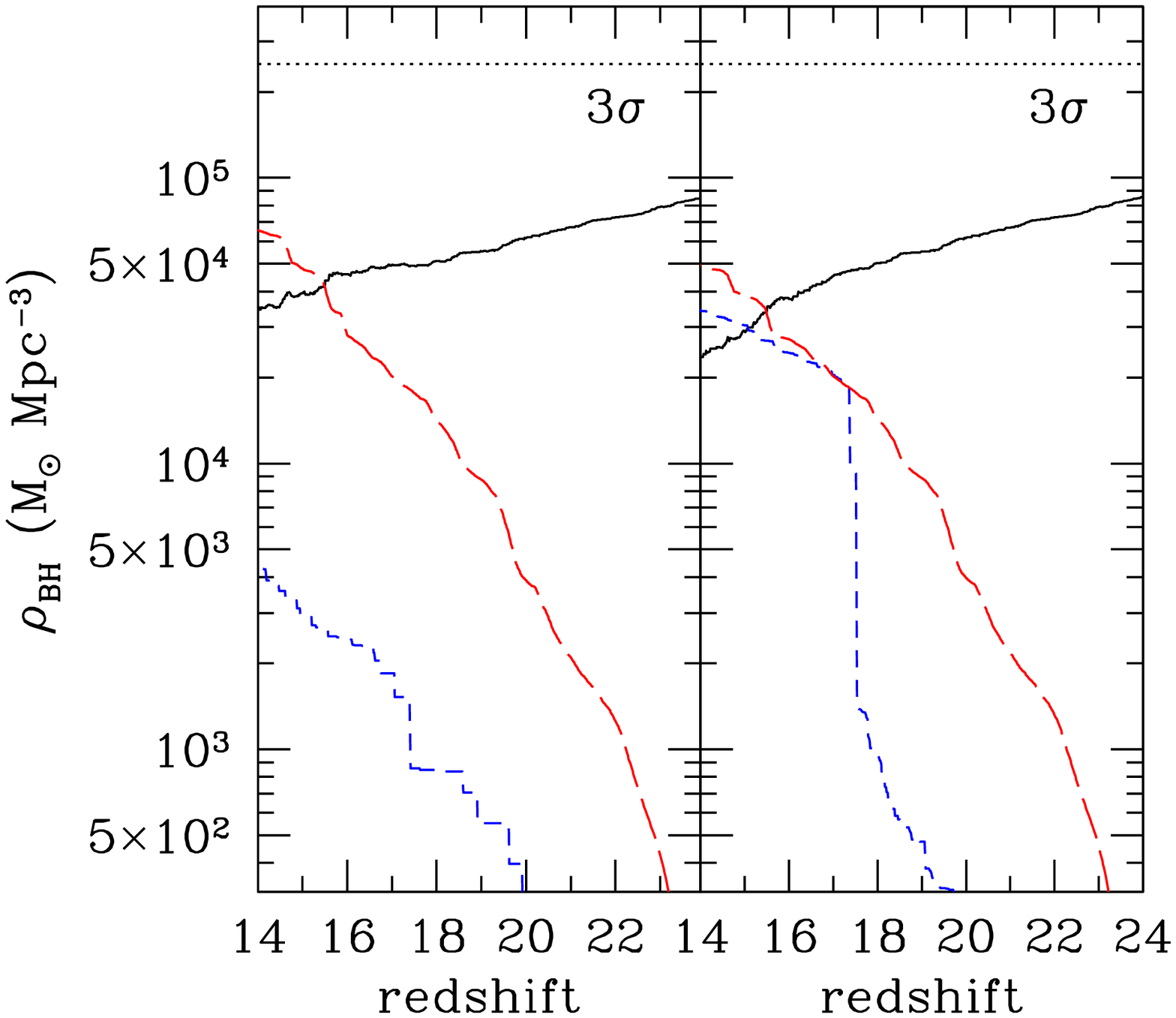,width=3.in}}
\caption{\footnotesize 
Contribution of nuclear, wandering, and intergalactic holes to the IMBH 
mass density as a function of redshift. In each major merger a mass
$\Delta m_{\rm acc}=10^{-5}\,M_h$ is accreted onto the BH in the main 
halo. {\it Left panel:} model in which (most) binaries stall. {\it Right panel:} 
model in which binaries shrink rapidly. {\it Solid line:} mass density 
of IMBHs in galaxy nuclei. {\it Long-dashed line:} wandering IMBHs 
retained in galaxy halos, mostly due to minor mergers. 
{\it Short-dashed line:} intergalactic IMBHs ejected from the host 
after a gravitational rocket. The horizontal dotted line shows the 
mass density of SMBHs in the
nuclei of nearby galaxies inferred by Yu \& Tremaine (2002).
}
\label{fig6}
\vspace{+0.5cm}
\end{figurehere}

Figure \ref{fig6} shows the predicted nuclear, wandering, and intergalactic 
IMBH mass density as a function of redshift, for the two sets of 
realizations described above. Two features are worth noting, the actual 
decrease with 
time of the density of nuclear IMBHs as more and more holes become `wandering'
after minor mergers (cf. Fig. \ref{fig2}), and the one order of magnitude 
increase in the density of intergalactic holes due to the gravitational 
rocket after rapid coalescence. In the case where (most) binaries stall, the 
small population (in mass density) of intergalactic IMBHs is due to a few
mergers involving near-equal mass light binaries in the most massive halos,
where the hardening timescale (eq. \ref{lnth}) is shorter.       
In the model where binaries coalesce rapidly we find recoil velocities 
that exceed the escape speed from their hosts 
in 80\% of the cases.  Indeed, because of the rocket effect, 
the mass density of intergalactic holes {\it exceeds} that in the 
nuclear variety below a redshift of 15 or so. 

\begin{figurehere}
\vspace{+0.2cm}
\centerline{
\psfig{file=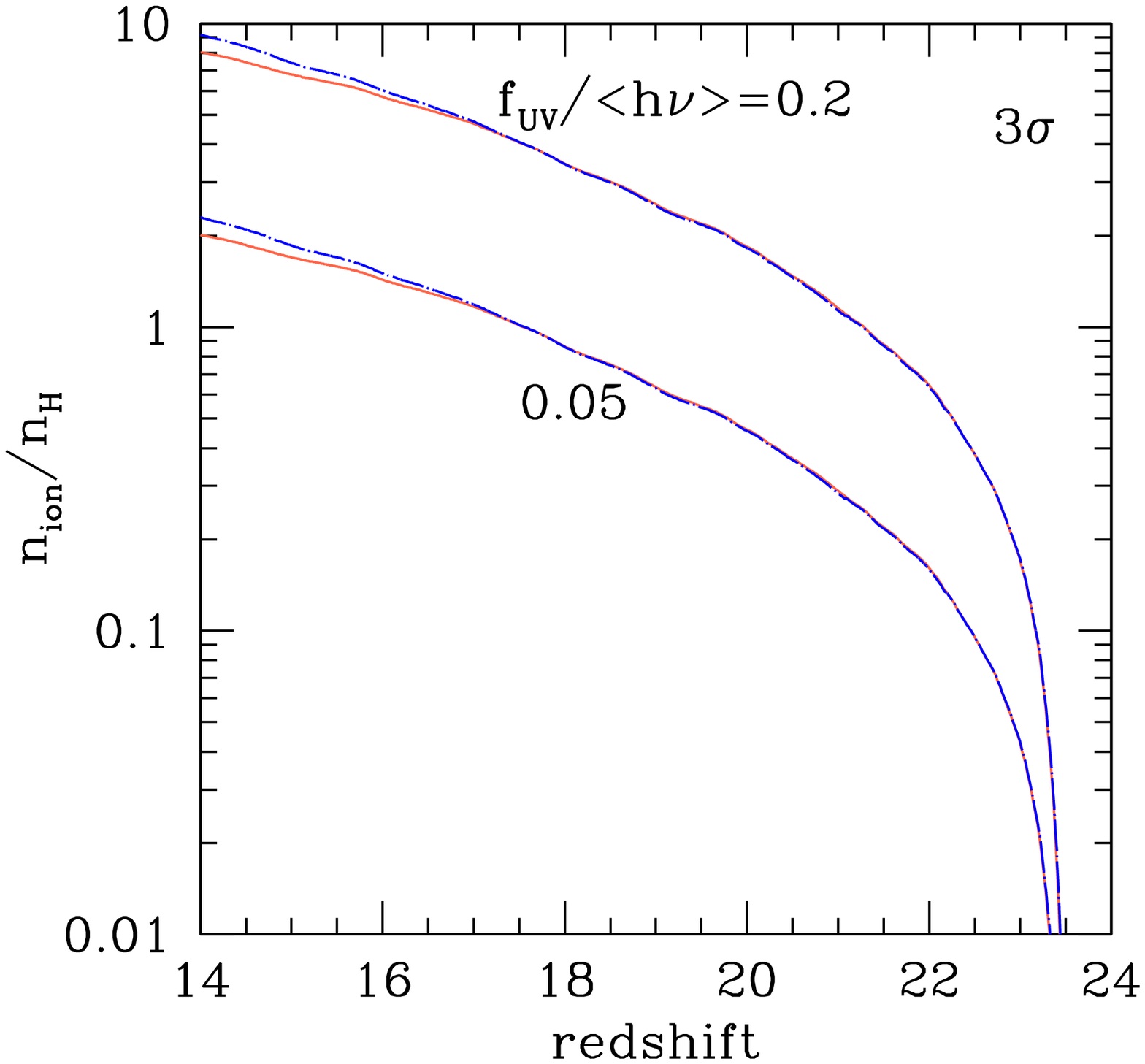,width=3.0in}}
\caption{\footnotesize Cumulative number of ionizing photons per 
hydrogen atom produced by miniquasars for different values of 
$f_{\rm UV}/\langle h\nu\rangle$ (in units of ryd$^{-1}$), assuming  
$\Delta m_{\rm acc}=10^{-5}\,M_h$. Seed IMBHs form
 in minihalos collapsing at $z=24$ from 3$\sigma$ density peaks.
{\it Dot-dashed curves:} model in which BH binaries stall.
{\it Solid curves:} same with rapid binary coalescence and gravitational
rocket.
}
\label{fig7}
\vspace{+0.5cm}
\end{figurehere}

The ejection into the IGM of IMBHs by the gravitational rocket effect 
appears -- depending on the initial density of seed holes and the ability 
of BH binaries to shrink and coalesce -- to decrease only slightly (by about
15\%) the `reionization efficiency' of miniquasars. This can be readily seen
in Figure \ref{fig7}, which shows the cumulative number of H-ionizing 
photons produced by miniquasars for the two 
cases (`binary stalling' vs. `instantaneous coalescence') discussed in this 
section. Even with a large reduction of the mass accreted by each
IMBH, the increased number of seeds produces more than a few ionizing
photons per hydrogen atom by redshift 15 for a spectrum with 
$f_{\rm UV}/\langle h\nu\rangle\gta 0.1\,$ryd$^{-1}$. If seed holes are 
as numerous as the 3$\sigma$ density peaks, the total number of 
Lyman-continuum photons produced per baryon by their stellar progenitors 
is now
\beq
{n_{\rm ion}\over n_b}\approx 70,000 f_*\approx 2.5,
\eeq 
i.e. sixty times larger than in the 3.5$\sigma$ case (cf. eq. \ref{ionb}). 
In this case miniquasars may still dominate the ionizing photon budget 
for $f_{\rm UV}/\langle h\nu\rangle\gta 0.2\,$ryd$^{-1}$
(or, of course, if $\Delta m_{\rm acc}>10^{-5}\,M_h$). 

\section{Reionization by hard photons}

We have seen above that, if cold gas in minihalos can be accreted either 
efficiently onto rare (`3.5$\sigma$') Population III IMBHs or much less 
efficiently
onto more numerous (`3$\sigma$') seed holes, then miniquasars may 
be responsible for cosmological reionization at 
early times. In this section we discuss the reheating and reionization 
history of an IGM photoionized primarily by a power-law metagalactic flux 
extending to X-ray energies, rather than by photons near the Lyman limit, 
as would be the case for purely stellar UV radiation. Reionization by 
energetic photons has been discussed recently by Venkatesan, Giroux, 
\& Shull (2001) and Oh (2001). 

\subsection{Soft X-ray background radiation}

Accretion onto IMBHs may be an attractive way to reionize the low-density
IGM. A large fraction of the UV radiation from massive 
stars may not escape the dense sites of star formation, or may be deposited   
locally in halo gas that recombines almost immediately. The harder radiation 
emitted from miniquasars is instead more likely to escape from the hosts
into intergalactic space, and may then produce more `durable' (albeit
partial) ionization in the diffuse IGM. 

In the pre-reionization universe, when the IGM is predominantly 
neutral, photons with energies above 24.6 eV will be absorbed as they
photoionize hydrogen or helium atoms. For a mixture of H and He with 
cosmic abundances, the effective bound-free absorption 
cross-section\footnote{The exact form of the photoionization cross-section
for \HI was taken from Spitzer (1978), and we have used the fit from 
Verner \etal (1996) for \HeI (see also Miralda-Escud\'e 2000).}\, can be 
approximated (to an accuracy of 30\% in the range $50\,{\rm eV}<
h\nu<2\,$keV) 
as $\sigma_{\rm bf}\approx 1.6\times 10^{-17}\,{\rm cm^2}~(h\nu/1\,{\rm 
ryd})^{-3}$. The continuum optical depth of a uniform
IGM of hydrogen density $n_0(1+z)^3$, for photons emitted at frequency 
$\nu_e$ by a source at redshift $z_e$ beyond the redshift of reionization 
$z_\rei$, is given by
\beq
\tau=\int_{z_\rei}^{z_e} dz\,{dl\over dz} n_0 (1+z)^3\,\sigma_{\rm bf}
(\nu)\approx 10^{6.4}\left({1\,{\rm ryd}\over h\nu_e}\right)^3,
\eeq
where $\nu\equiv \nu_e(1+z)/(1+z_e)$, $dl/dz\simeq H_0^{-1}\Omega_M^{-1/2}
(1+z)^{-5/2}$, and we have evaluated the integral from $z_\rei=14$
to $z_e=19$.
The universe is optically thick even to hard radiation, and all photons 
below $(h\nu)_{\rm thick}\sim 2\,$keV will be absorbed across a Hubble volume. 
Since a 
spectrum with $\nu\,L_\nu\sim$ const (like the nonthermal component observed 
in ULXs) is characterized by equal power per logarithmic frequency interval, 
photoelectric absorption by the host galaxies will not significantly 
attenuate the ionizing energy flux.

The mean free path of ionizing photons in the neutral IGM is 
\beq 
\lambda={1\over n_0(1+z)^3\,\sigma_{\rm bf}}\approx 13\,{\rm pc}\,
\left({1+z\over 20}\right)^{-3}\,\left({h\nu\over 1\,{\rm ryd}}\right)^3.
\eeq
At $z=24$, the comoving space density of halos collapsing from 3.5$\sigma$
density peaks is $\sim 15\,$Mpc$^{-3}$, corresponding to a mean
proper distance between neighboring halos of $15\,$kpc. Photons 
with energies above $(h\nu)_{\rm overlap} \sim 150$ eV will then have a mean 
free path greater than the mean separation between sources. We can therefore 
separate the radiation field into two components: a {\it fluctuating EUV field} 
in the range $13.6<h\nu<150\,$eV, $J_{\rm UV}$, which creates expanding 
patchy \HII regions, and a {\it nearly uniform soft X-ray background} 
component in the range $150\,{\rm eV}<h\nu<2\, {\rm keV}$, $J_X$, 
which partially ionizes the IGM in a fairly homogeneous manner. 

\subsection{Secondary ionizations}

Prior to the epoch of reionization breakthrough, before the overlapping of 
fully ionized (by EUV photons) bubbles around individual miniquasars, 
soft X-ray radiation will produce a warm ($T_{\rm IGM}\sim\,$ few $\,\times 
100-1000\,$K), weakly ionized ($x\sim 0.1$) IGM (Venkatesan \etal 2001).
X-rays alone do not produce a fully ionized medium, but can partially
photoionize the gas by repeated secondary ionizations.
A primary nonthermal photoelectron of energy $E=1\,$keV in 
a medium with residual ionization (from the recombination epoch) 
$x=2\times 10^{-4}$ will create over two dozens secondary electrons, 
depositing a fraction $f_{\rm ion}\approx 37\%$ of its initial energy as 
secondary ionizations of hydrogen, and only $f_{\rm heat}\approx 13\%$ as
heat (Shull \& Van Steenberg 1985). 

\begin{figurehere}
\vspace{+0.2cm}
\centerline{
\psfig{file=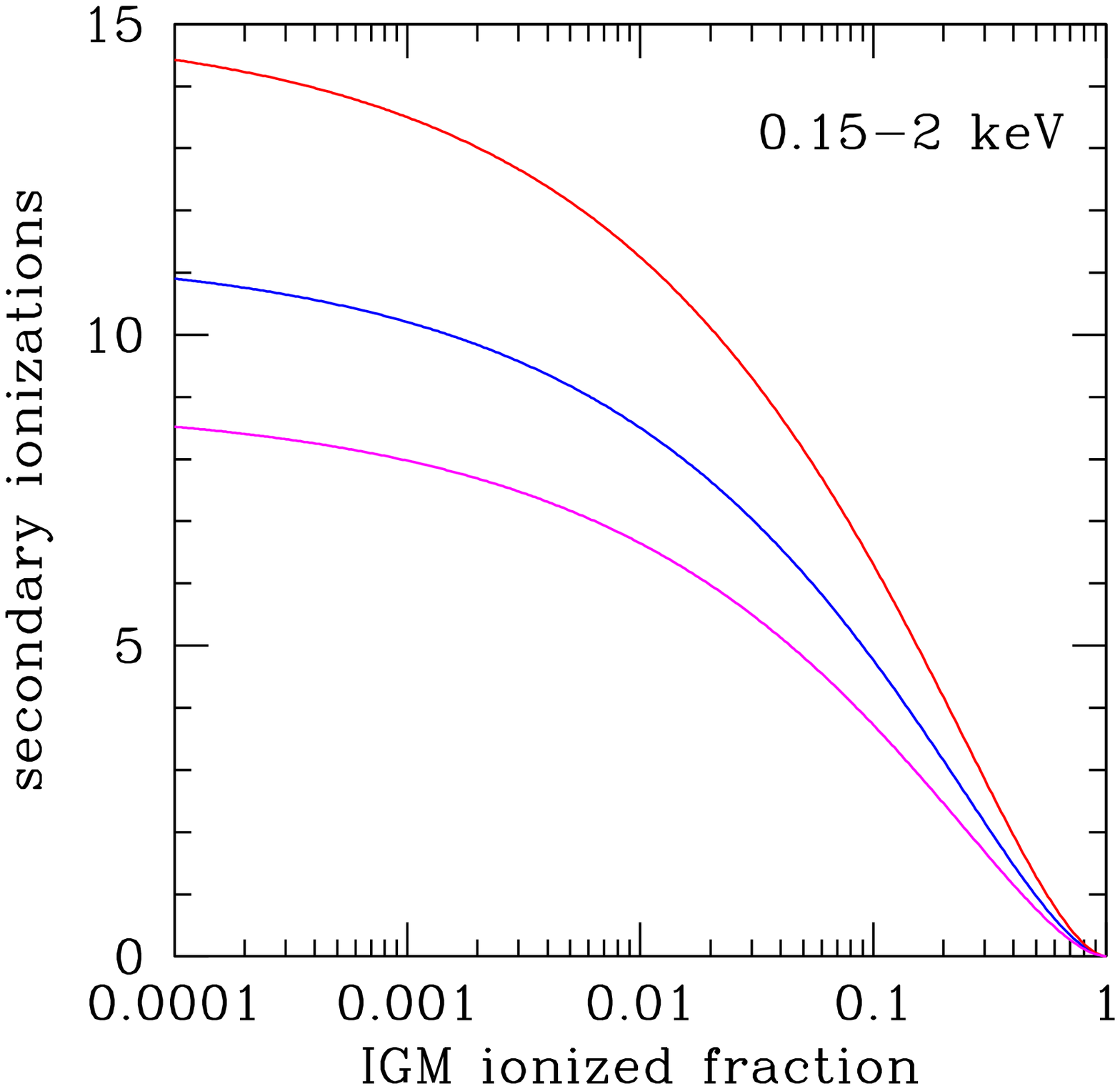,width=3.0in}}
\caption{\footnotesize Number of H secondary ionizations per ionizing photon 
vs. IGM ionized fraction. A power-law radiation field with energy 
spectrum $J_X\propto \nu^{-\alpha}$, extending from 150 eV to 2 keV, has 
been assumed. The three curves from top to bottom are for $\alpha=0.5, 1, 
1.5$, respectively. As the medium becomes more ionized an increasing 
fraction of energy is deposited as heat. 
}
\label{fig8}
\vspace{+0.5cm}
\end{figurehere}

Figure \ref{fig8} shows the mean number
of secondary ionizations per ionizing photon as a function of IGM ionized 
fraction, for a power-law spectrum extending from 150 eV to 2 keV (`uniform soft
X-ray background'). The calculation uses the analytic fits given by Shull \& Van 
Steenberg (1985). The timescale for electron-electron encounters resulting 
in a fractional energy loss $f=\Delta E/E$, 
\beq 
t_{\rm ee}\approx 140\,{\rm yr}~Ef\,\left({1+z\over 20}\right)^{-3}\,
\left({\ln\Lambda\over 20}\right)^{-1}\,x^{-1}
\eeq
(where $E$ is measured in keV), is typically much shorter that the 
electron Compton 
cooling timescale off cosmic microwave background (CMB) photons, 
$t_C=(7\times 10^6\, {\rm yr})\,[(1+z)/20]^{-4}$, 
and thus the primary photoelectron will 
ionize and heat the surrounding medium before it is cooled by the CMB.
 
The history of reheating and reionization by a hard $\nu^{-1}$ spectrum may 
then be described as follows. Initially, when the IGM is mostly neutral,
each photoionization produces a host of secondary collisional ionizations,
with about one hydrogen secondary ionization for every 37 eV of energy in the 
primary photoelectron.
For photon energies above 0.2 keV, the ratio of the  \HI/\HeI photoionization 
cross-sections drops below 4\%: since the 
primordial ratio of helium to hydrogen is about 8\%, the photoionization
and heating rates will be dominated by helium absorption, which exceeds the 
hydrogen contribution by $\sim 2:1$. While \HeI is the main source of hot 
primary photoelectrons, however, \HI undergoes the bulk of secondary 
ionizations. Secondary ionizations operate efficiently for \HI and \HeI,
but not for \HeII (Shull \& Van Steenberg 1985). Because of
losses to secondary ionizations and to collisional excitations of \HI
and \HeI before thermalization, the heating by energetic photons
is initially not as efficient as for photons near the photoelectric
thresholds. The amount of heat deposited 
into a unit proper volume of the IGM by the mean field can be written in 
first approximation as 
\beq
\Delta Q=f_{\rm heat}{4\pi\over c}\int_{\nu_{\rm overlap}}^{\nu_{\rm thick}}
d\nu J_X= (1+z)^3f_{\rm heat}f_X\epsilon\rho_{\rm acc}c^2,
\eeq
where $f_X$ is the fraction of the bolometric emissivity of miniquasars 
that is radiated between $\nu_{\rm overlap}$ and $\nu_{\rm thick}$,
$f_X=\ln(\nu_{\rm thick}/\nu_{\rm overlap})/\ln(\nu_2/\nu_1)$
for a $\nu^{-1}$ spectrum extending from $\nu_1$ to $\nu_2$. For the spectrum 
we have chosen, the result is only logarithmically sensitive to the 
integration bounds, and we set $f_X=0.4$. In the scenarios for the formation
and growth of IMBHs discussed in the previous sections, gas accretion 
onto IMBHs rapidly build up (by $z\sim 22$) a comoving mass density 
$\rho_{\rm acc}$ in excess of a few hundred $\mden$, enough to heat 
up the gas to initial temperatures:
\beq
T={\Delta Q\over 3/2nk}\approx  1500\,{\rm K}\,\left({f_{\rm heat}\over 
0.13}\right)\left({\rho_{\rm acc}\over 200\,\mden}\right).
\eeq
This same radiation field would ionize the IGM to: $x\approx 4\times 
10^{-3}\,(f_{\rm ion}/0.37)\,(\rho_{\rm acc}/200\,\mden)$.
Since both the hydrogen recombination and the Compton cooling
time are much longer than the then Hubble time for $x<0.1$, the 
gas evolves adiabatically. 
Once the IGM ionized fraction increases to $x\approx 0.1$, 
the number of secondary ionizations per ionizing photon drops to a 
few (Fig. \ref{fig8}), and the bulk of the primary's energy goes 
into heat ($f_{\rm heat}\approx 0.6$) via elastic Coulomb collisions 
with thermal electrons. At this point Compton cooling will restrict 
any further increase in temperature: it is the rate of direct 
photoionizations by UV photons in the soft end of the spectrum that drives 
the final stages of the reionization process, until the fully ionized 
bubbles associated with individual sources finally overlap.   

As \HI and \HeI reionization nears completion, the only means for
a hard radiation field to inject energy into the plasma is via \HeII
photoionization and Compton heating. Compton heating is much less
efficient than \HeII photoionization heating: the small Thompson 
cross-section implies that the universe is always optically thin to 
Compton heating of electrons. Furthermore, the energy exchange per 
interaction is only a small fraction of the photon energy, 
$\sim h\nu (h\nu/m_ec^2)$. Thus, in most situations, Compton heating
will play a sub-dominant role compared to \HeII photoheating 
(cf. Madau \& Efstathiou 1999). 

\subsection{Entropy floor}

As recently pointed out by Oh \& Haiman (2003), soft X-rays can exert 
another important effect on the IGM and subsequent star formation, as 
the warm, partially ionized medium produced by energetic photons has
a high entropy. Gas at the mean density that is heated to temperatures
\begin{equation}
T_{\rm IGM} > 90\,{\rm K} \left( \frac{T_{\rm vir}}{3000\, {\rm K}} \right)
\left( \frac{\delta}{200} \right)^{-2/3}
\end{equation}
(where $\delta$ is the overdensity of the gas in the minihalo in the
absence of preheating) will have entropy in excess of that acquired by
gravitational shock heating alone. It will therefore have excess
pressure after adiabatic heating and compression. As shown in the 
previous section, such temperatures are easily achievable with miniquasars.
This entropy floor: (a) greatly reduces gas clumping, curtailing the 
number of photons needed to maintain reionization; and (b) results 
in significantly lower gas densities in the cores of minihalos that 
suppress rapid ${\rm H_{2}}$
formation. The latter effect may imply that X-rays inhibit rather than 
enhance star formation (Oh \& Haiman 2003).

An entropy floor could conceivably also reduce the efficiency with
which miniquasars can shine. In halos that have had their central densities 
reduced and are unable to form stars, IMBHs may still accrete gas, but at 
a considerably reduced rate. Miniquasars, however, are $\sim 15-150$ times 
more efficient at producing ionizing radiation per processed baryon 
than metal-free stars, and thus need a much smaller reservoir of cold gas 
to reionize the universe. If $f_{UV}/\langle h\nu \rangle \sim 0.2$, and 
$100\,$MeV are released per baryon, then only
$\sim 7 \times 10^{-7}$ of all baryons need to be accreted onto IMBHs 
to provide one ionizing photon per baryon, compared to $\sim 10^{-5}$ for
massive metal-free stars (Bromm, Kudritzki, \& Loeb 2001), and $\sim
10^{-4}$ for metal-free stars with a Salpeter IMF (Tumlinson \& Shull
2000). The cold gas that cooled {\it prior} to the establishment
of the entropy floor (and formed an accretion disk around the central
black hole) will still be available as fuel. If only a fraction
$f_{\rm cool}\sim 1\%$ of the gas in a $3.5\sigma$ peak halo cooled 
to form an accretion 
disk around the central black hole before an entropy floor was established, 
this still represents $\sim 5 \times 10^{-6} (f_{\rm cool}/0.01)$ of all
baryons, enough to provide 10 ionizing photons per hydrogen atom; 
this number increases by an order of magnitude if 3$\sigma$ peaks are
involved instead. By contrast, these baryon fractions are insufficient to
reionize the universe with Population III metal-free stars; the entropy floor
would prohibit full reionization by such objects. The combination
of X-rays (which reduce gas clumping, and thus the number of UV
photons require to complete reionization) and UV photons (which carry
reionization to completion) could allow miniquasars to reionize the
universe quite efficiently.   

\section{Summary}

This paper should be read as a first attempt at treating the impact on 
the very early IGM of Population III IMBHs -- the remnants of the first 
generation of massive stars -- in the 
context of hierarchical structure formation theories. We have tried to 
incorporate some of the essential astrophysics into a scenario  
for the reheating and reionization of the universe by miniquasars.
In our model quasar activity is driven by major 
mergers and IMBHs are able to accrete at the Eddington rate only in 
the densest 
inner regions of the merger remnant. If seed IMBHs are as rare as the 
3.5$\sigma$ peaks of the primordial density field, they evolve largely in 
isolation; a significant number of BH binary systems forms if IMBHs 
populate the more numerous 3$\sigma$ peaks instead. In the case of rapid 
binary coalescence, rather than accrete and shine as miniquasars, a 
fraction of IMBHs will be displaced from galaxy centers and ejected into the 
IGM by the `gravitational rocket' effect. Note that, for gravitational 
radiation to induce coalescence, a hard BH binary must decay by a larger 
factor in minihalos with small velocity dispersion than in more massive 
galaxies. The loss of orbital angular momentum to a gaseous disk, rather 
than via three-body interactions with DM particles or stars, may drive
IMBH binaries to merge rapidly, but the details are unclear.     

Under a number of different assumptions 
for the amount of gas accreted onto IMBHs and their emission spectrum, 
miniquasars powered by IMBHs -- and not their metal-free stellar 
progenitors -- may be responsible for cosmological reionization at 
$z\sim 15$. The {\it WMAP} data on $\tau_e$ 
may then be setting constraints on gas accretion onto the earliest 
generation of IMBHs, as well as on the IMF of the first burst of star 
formation in the universe. Even if the `UV radiation efficiency' of miniquasars 
was lower and Population III stars were the dominant 
source of ionizing radiation at $z\sim 25$, the transition from Population III to 
Population II at a redshift possibly determined by the level of metal 
enrichment of the IGM would 
decrease by more than an order of magnitude the emission rate of 
Lyman-continuum photons (e.g. Cen 2003b). Hard radiation from miniquasars 
could then sustain ionization in the redshift range $z_\rei<z<20$ (say), 
before ordinary stars with a Salpeter IMF took over. 

At $z\gta 4-5$, the known population of quasars appears not to 
contribute significantly to the hydrogen ionizing background (Madau,
Haardt, \& Rees 1999). The comoving space density of luminous SDSS quasars
at $z\sim 6$ is 20 times smaller than that at $z\sim 3$ (Fan \etal 2003).
The dearth of bright QSOs at $z\sim 6$ may be reconciled with
the existence of a substantial population of accreting IMBHs at $z\sim 15$ if there 
were two distinct periods of quasar activity. The first may have ended with the demise
of miniquasars due to radiative feedback 
mechanisms after the reionization epoch (as photoionization from background 
UV photons unbinds the gas in the shallow potential wells of minihalos,
Barkana \& Loeb 1999; Haiman, Abel, \& Madau 2001) or to a raised entropy floor. 
Efficient gas accretion onto 
nuclear BHs could conceivably start again only at $z\lta 10<z_\rei$, in $T_{\rm vir}\gg 
10^{4}\,$K halos further down in the merger hierarchy that are not 
susceptible to Jeans smoothing and gas cooling suppression effects. 
This may lead to a second ``quasar era" seeded by more massive 
($\sim 5000\,\msun$) pre-existing holes.  

One might worry that with a sufficiently
hard ionizing spectrum \HI and \bHeII$\rightarrow$\HeIII 
reionization may be simultaneous,
whereas observations of the \HeII \Lya\ forest have been interpreted as 
evidence that the double reionization of helium occurred at a redshift 
of $\sim 3$ (e.g. Kriss et al. 2001; Reimers \etal 1997). For a 
$\nu^{-1}$ miniquasar spectrum there are 3 times 
more \HeII ionizing photons ($h\nu>4\,$ryd) per He atom than photons above 
1 ryd per H atom.
As doubly ionized helium recombines about 6 times faster than hydrogen, 
the number of $>4\,$ ryd photons per He atom emitted in one recombination 
timescale is only half of the corresponding value for hydrogen, and
there will be just a small delay between the complete overlapping of 
\HII and \HeIII regions (Miralda-Escud\`e \& Rees 1993; Madau \& Meiksin 
1994). \HeII reionization by miniquasars at $z\gta 10$ would provide a 
significant boost to the gas temperature, with interesting implications 
for the thermal history of the IGM (e.g. Hui \& Haiman 2003).
The patchy \HeII absorption observed at $z\sim 3$ could then be explained
by variations in the spectra of the ionizing sources (Smette \etal 2002),
or may imply that \HeII was reionized twice: a first time by 
miniquasars, followed by recombination as Population II stars with
softer spectra took over, and then by a second overlap phase at $z\sim 3$ during the second 
quasar epoch.

Absorption in the intervening IGM
will make miniquasars at $z\sim 15$ inaccessible to direct observations
from 2$\,\mu$m down to soft X-ray energies.  \HeII recombination lines longward of 
\HI \Lya\ may be potentially detectable by the {\it James Webb Space Telescope}
only from the brightest miniquasars (Oh, Haiman, \& Rees 2001).
An alternative way to probe the end of the cosmic `dark ages' and 
discriminate between different reionization histories is through 21 cm 
tomography of neutral hydrogen (Madau, Meiksin, \& Rees 1997). In general, 
21 cm spectral 
features will display angular structure as well as structure in redshift 
space due to inhomogeneities in the gas density field, hydrogen ionized 
fraction, and spin temperature. As recently discussed by Tozzi \etal (2000), Ciardi 
\& Madau (2003), and Furlanetto, Sokasian, \& Hernquist (2003), the fluctuations 
in 21 cm line induced by 
the `cosmic web' that develops at early times in CDM-dominated cosmologies
could be detected {\it in emission} against the CMB provided that: (a) 
at epochs when the IGM is still mainly neutral, the first generation of
(Population III) stars produce enough UV continuum photons with energies 
between 10.2 and 13.6 eV to mix the hyperfine levels and decouple the 
hydrogen spin temperature from $T_{\rm CMB}$; and (b) the IGM is `warm', 
i.e. $T_{\rm IGM}\gg T_{\rm CMB}=57\,$K $[(1+z)/20]$. X-ray radiation 
naturally produces a warm, weakly ionized IGM. Miniquasars turning
on at early stages may then make the IGM visible in 21 cm emission 
as structure develops, before the universe is actually reionized.
Brightness temperature fluctuations may be detectable by future facilities
like the {\it LOw Frequency ARray} ({\it LOFAR}) if foreground contamination 
from unresolved extragalactic point sources  (Di Matteo \etal 2002; Oh 
\& Mack 2003) can be successfully removed by using spectral structure in 
frequency space.

Finally, we remark that accreting IMBHs may release large amounts of energy 
to their environment in other forms than radiative output. Outflows from
miniquasars may inject kinetic energy into intergalactic space and 
raise the temperature of the IGM to a much higher adiabat than expected
from photoionization. In the type of scenario 
discussed in \S~3, gas accretion along cosmic history builds up 
by $z\sim 14$ a comoving mass density, $\rho_{\rm acc}$, in excess of 
$10^4\,\mden$. If a fraction $f_w$ of the accreted rest-mass energy
was used to drive an outflow and ultimately deposited into the IGM, 
the energy input per baryon at this epoch,
\beq
E_w=f_w\,{\rho_{\rm acc}c^2\over n_b}\sim 0.1\,{\rm keV}
\left({f_w\over 0.05}\right)\,\left({\rho_{\rm acc}\over 10^4\,\mden}\right),
\eeq
would suffice to preheat vast regions of the universe to temperatures
above a few $\times 10^5\,$K, and so inhibit the formation of early dwarf 
galaxies (Benson \& Madau 2003). The detailed thermal history of the 
IGM may then depend on different forms of `feedback' mechanisms from 
miniquasars during the epoch of galaxy formation.  

\acknowledgements
Support for this work was provided by NASA grant NAG5-11513 and NSF 
grant AST-0205738 (P.M.), by grant MIUR COFIN2002 (F.H.), and 
by the Royal Society (M.J.R.).
 
{}

\end{document}